\documentstyle[graphicx,aps,multicol]{revtex} 

\draft

\begin{document}

\title{Critical temperature for quenching of pair correlations}
\author{A.~Schiller, A.~Bjerve, M.~Guttormsen, M.~Hjorth-Jensen, 
F.~Ingebretsen, E.~Melby, S.~Messelt, J.~Rekstad, S.~Siem, and 
S.W.~{\O}deg{\aa}rd}
\address{Department of Physics, University of Oslo, 
P.O.Box 1048 Blindern, N-0316 Oslo, Norway}

\maketitle

\begin{abstract}
The level density at low spin in the $^{161,162}$Dy and $^{171,172}$Yb nuclei 
has been extracted from primary $\gamma$ rays. The nuclear heat capacity is 
deduced within the framework of the canonical ensemble. The heat capacity  
exhibits an S-formed shape as a function of temperature, which is interpreted
as a fingerprint of the phase transition from a strongly correlated to an 
uncorrelated phase. The critical temperature for the quenching of pair 
correlations is found at $T_c=0.50(4)$~MeV. 
\end{abstract}


\begin{multicols}{2}

The thermodynamical properties of nuclei deviate from infinite systems. 
While the quenching of pairing in superconductors is well described as a 
function of temperature, the nucleus represents a finite many body system 
characterized by large fluctuations in the thermodynamic observables. A 
long-standing problem in experimental nuclear physics has been to observe the 
transition from strongly paired states, at around $T=0$, to unpaired states at 
higher temperatures. 

In nuclear theory, the pairing gap parameter $\Delta$ can be studied as 
function of temperature using the BCS gap equations \cite{SY63,Go81}. From this
simple model the gap decreases monotonically to zero at a critical temperature 
of $T_c\sim 0.5\,\Delta$. However, if particle number is projected out 
\cite{FS76,DK95}, the decrease is significantly delayed. The predicted decrease
of pair correlations takes place over several MeV of excitation energy 
\cite{DK95}. Recently \cite{MB99}, we reported structures in the level 
densities in the 1--7~MeV region, which are probably due to the breaking of 
nucleon pairs and a gradual decrease of pair correlations. 

Experimental data on the quenching of pair correlations are important as a 
test for nuclear theories. Within finite temperature BCS and RPA models, level 
density and specific heat are calculated for e.g.\ $^{58}$Ni \cite{Ng90}; 
within the shell model Monte Carlo method (SMMC) \cite{LJ93,KD97} one is now 
able to estimate level densities \cite{Or97} in heavy nuclei \cite{WK98} up to 
high excitation energies. 

The subject of this Letter is to report on the observation of the gradual 
transition from strongly paired states to unpaired states in rare earth nuclei 
at low spin. The canonical heat capacity is used as a thermometer. Since only 
particles at the Fermi surface contribute to this quantity, it is very 
sensitive to phase transitions. It has been demonstrated from SMMC calculations
in the Fe region \cite{RH98,AL99,Al99}, that breaking of only one nucleon pair 
increases the heat capacity significantly. 

Recently \cite{HB95,SB99}, we presented a method for extracting level density
and $\gamma$ strength function from measured $\gamma$ ray spectra. Since the 
$\gamma$ decay half lives are long, typically $10^{-12}$--$10^{-19}$~s, the 
method should essentially give observables from a thermalized system 
\cite{SG99}. The spin window is typically 2--8~$\hbar$ and the excitation 
energy resolution is 0.3~MeV.

The experiments were carried out with 45~MeV $^3$He projectiles from the MC-35 
cyclotron at the University of Oslo. The experimental data were recorded with 
the CACTUS multidetector array \cite{GA90} using the ($^3$He,$\alpha\gamma$) 
reaction on $^{162,163}$Dy and $^{172,173}$Yb self-supporting targets. The beam
time was two weeks for each target. The charged ejectiles were detected with 
eight particle telescopes placed at an angle of 45$^\circ$ relative to the beam
direction. Each telescope comprises one Si~$\Delta E$ front and one Si(Li)~$E$ 
end detector with thicknesses of 140 and 3000~$\mu$m, respectively. An array of
28~5"$\times$5"~NaI(Tl) $\gamma$ detectors with a total efficiency of 
$\sim$15\% surrounded the target and particle detectors. 

From the reaction kinematics the measured $\alpha$ particle energy can be 
transformed to excitation energy $E$. Thus, each coincident $\gamma$ ray can be
assigned a $\gamma$ cascade originating from a specific excitation energy. The 
data are sorted into a matrix of $(E,E_{\gamma})$ energy pairs. At each 
excitation energy $E$ the NaI $\gamma$ ray spectra are unfolded \cite{GT96}, 
and this matrix is used to extract the primary $\gamma$ ray matrix, with the 
well established subtraction technique of Ref.~\cite{GR87}. 

The resulting matrix $P(E,E_{\gamma})$, which describes the primary 
$\gamma$ spectra obtained at an initial excitation energy $E$, is factorized 
according to the Brink-Axel hypothesis \cite{Br55,Ax62} by 
$P(E,E_\gamma)\propto\rho(E-E_\gamma)\sigma(E_\gamma)$. The level density
$\rho(E)$ and the $\gamma$ strength function $\sigma(E_\gamma)$ are determined
by a least $\chi^2$ fit to $P$. Since the fit yields an infinitely large number
of equally good solutions, which can be obtained by transforming one arbitrary 
solution by 
\begin{eqnarray}
\tilde{\rho}(E-E_\gamma)&=&A\exp[\alpha(E-E_\gamma)]\rho(E-E_\gamma),\\
\label{eq:transformation}
\tilde{\sigma}(E_\gamma)&=&B\exp(\alpha E_\gamma)\sigma(E_\gamma),
\end{eqnarray}
we have to determine the parameters $A$, $B$ and $\alpha$ by comparing the 
$\rho$ and $\sigma$ functions to known data. The $A$ and $\alpha$ parameters 
are fitted to reproduce the number of known levels in the vicinity of the 
ground state \cite{FS96} and the neutron resonance spacing at the neutron 
binding energy \cite{LH75,MC70,LC73}. 

\begin{figure}\centering
\includegraphics[totalheight=17.9cm]{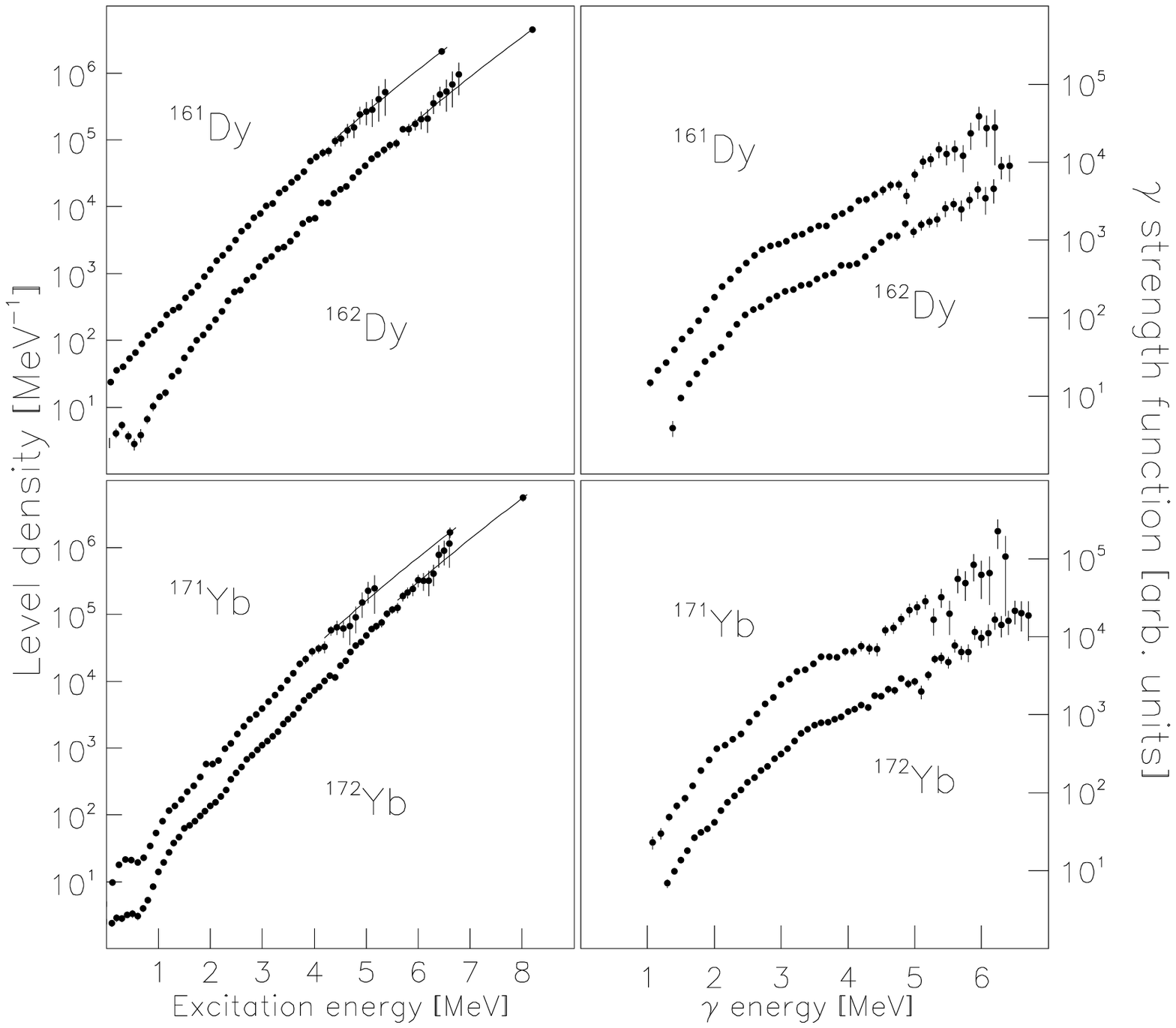}
\caption{Experimental level density (points in left panels) and $\gamma$ 
strength function (right panels) for $^{161,162}$Dy (upper part) and 
$^{171,172}$Yb (lower part). The error bars show the statistical uncertainties.
The solid lines are extrapolations based on a shifted Fermi gas model (see 
text). The isolated point at the neutron binding energy is obtained from 
neutron resonance spacing data.}
\label{fig:leveldensity}
\end{figure}

Figure~\ref{fig:leveldensity} shows the extracted level densities and $\gamma$
strength functions for the $^{161,162}$Dy and $^{171,172}$Yb nuclei. The data 
for the even nuclei are published recently \cite{MB99}, and are included to 
be compared to odd nuclei. Also the $\gamma$ strength functions are given for 
all four nuclei. Since other experimental data on the $\gamma$ strength 
function in the energy interval 1.5--6~MeV is sparse, we could not fix the 
parameter $B$; and therefore we give the functions in arbitrary units. However,
when fitting the functions with the expression $C\,E_\gamma^n$ we obtain 
$n\sim 3.9,\,4.5$ for the Dy and Yb isotopes respectively, which one would 
expect from the tale of the giant dipole resonance \cite{HB95}. In the 
following, only the level density will be discussed. 

The partition function in the canonical ensemble 
\begin{equation}
Z(T)=\sum_{n=0}^\infty\rho(E_n)e^{-E_n/T} 
\end{equation}
is determined by the measured level density of accessible states $\rho(E_n)$ in
the present nuclear reaction. Strictly, the sum should run from zero to 
infinity. In this work we calculate $Z$ for temperatures up to $T=1$~MeV. 
Assuming a Bethe-like level density expression \cite{Be36}, the average 
excitation energy in the canonical ensemble
\begin{equation}
\langle E(T)\rangle=Z^{-1}\sum_{n=0}^{\infty}E_n\rho(E_n)e^{-E_n/T}
\label{eq:energy}
\end{equation}
gives roughly $\langle E\rangle\sim aT^2$ with a standard deviation of 
$\sigma_E\sim T\sqrt{2aT}$ where $a$ is the level density parameter. Using 
Eq.~(\ref{eq:energy}) requires that the level density should be known up to 
$\langle E\rangle+3\sigma_E$, typically 40~MeV. However, the experimental level
densities of Fig.~\ref{fig:leveldensity} only cover the excitation region up 
close to the neutron binding energy of about 6 and 8~MeV for odd and even mass 
nuclei, respectively. For higher energies it is reasonable to assume Fermi gas 
properties, since single particles are excited into the continuum region with 
high level density. Therefore, due to lack of experimental data, the level 
density is extrapolated to higher energies by the shifted Fermi gas model 
expression \cite{GC65} 
\begin{equation}
\rho_{\mathrm{FG}}(U)=f\,\frac{\exp(2\sqrt{aU})}
{12\,\sqrt{0.1776}\,a^{1/2}\,U^{3/2}\,A^{1/3}}, 
\end{equation}
where $U$ is the shifted energy and $A$ is the mass number. For shift and level
density parameters $a$, we use von~Egidy's parameterization of this expression 
\cite{ES87}. The expression had to be normalized by a factor $f$ in order to 
match the neutron resonance spacing data. The factors are 1.2, 0.94, 0.4 and 
0.6 for $^{161,162}$Dy and $^{171,172}$Yb respectively. The solid lines in 
Fig.~\ref{fig:leveldensity} show how the expression extrapolates our 
experimental level density curves. 

The extraction of the microcanonical heat capacity $C_V(E)$ gives large 
fluctuations which are difficult to interpret \cite{MB99}. Therefore, the heat 
capacity $C_V(T)$ is calculated within the canonical ensemble, where $T$ is a 
fixed input value in the theory, and a more appropriate parameter. The heat 
capacity is then given by 
\begin{equation}
C_V(T)=\frac{\partial\langle E\rangle}{\partial T},
\end{equation}
and the averaging made in Eq.~(\ref{eq:energy}) gives a smooth temperature 
dependence of $C_V(T)$. A corresponding $C_V(\langle E\rangle)$ may also be 
derived in the canonical ensemble.

\begin{figure}\centering
\includegraphics[totalheight=21.5cm]{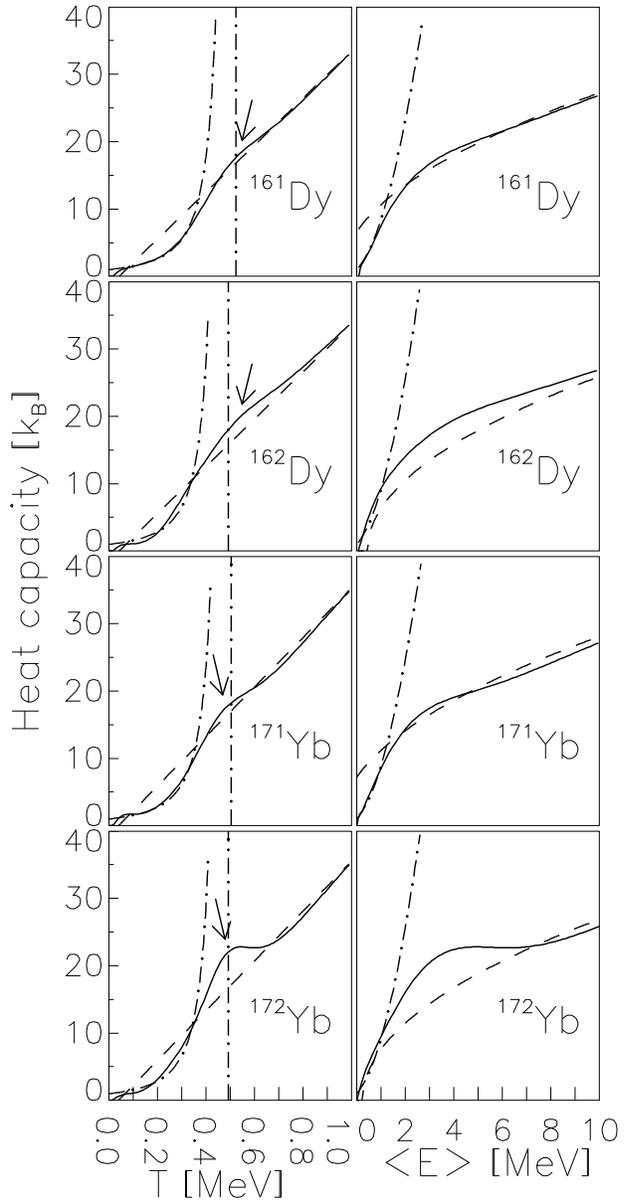}
\caption{Semi-experimental heat capacity as function of temperature (left 
panels) and energy $\langle E\rangle$ (right panels) in the canonical ensemble 
for $^{161,162}$Dy and $^{171,172}$Yb. The dashed lines describe the 
approximate Fermi gas heat capacity. The arrows indicate the first local 
maxima of the experimental curve relative to the Fermi gas estimates. The 
dashed dotted lines describe estimates according to Eq.~(\protect{\ref{eq:Tc}})
where $\tau$ is set equal to the critical temperature $T_c$. $T_c$ is indicated
by the vertical lines.}
\label{fig:heatcapacity} 
\end{figure}

The deduced heat capacities for the $^{161,162}$Dy and $^{171,172}$Yb nuclei 
are shown in Fig.~\ref{fig:heatcapacity}. All four nuclei exhibit similarly 
S-shaped $C_V(T)$-curves with a local maximum relative to the Fermi gas 
estimate at $T_c\approx 0.5$~MeV. The S-shaped curve is interpreted as a 
fingerprint of a phase transition in a finite system from a phase with strong 
pairing correlations to a phase without such correlations. Due to the strong 
smoothing introduced by the transformation to the canonical ensemble, we do not
expect to see discrete transitions between the various quasiparticle regimes, 
but only the transition where all pairing correlations are quenched as a whole.
In the right panels of Fig.~\ref{fig:heatcapacity}, we see that 
$C_V(\langle E\rangle)$ has an excess in the heat capacity distributed over 
a broad region of excitation energy and is not giving a clear signal for 
quenching of pairing correlations at a certain energy \cite{MB99}. 

In order to extract a critical temperature for the quenching of pairing 
correlations from our data, we have to be careful, not to depend too much on 
the extrapolation of $\rho$. An inspection of Fig.~\ref{fig:leveldensity} shows
that the level density is roughly composed of two components as proposed by 
Gilbert and Cameron \cite{GC65}: (i) a low energetic part; approximately a 
straight line in the log plot, and (ii) a high energetic part including the 
theoretical Fermi gas extrapolation; a slower growing function. For 
illustration, we construct a simple level density formula composed of a 
constant temperature level density part with $\tau$ as temperature parameter, 
and a Fermi gas expression 
\begin{equation}
\rho(E)\propto\left\{\begin{array}{cc}\eta\exp(E/\tau)&{\mathrm{for}}\ E\leq
\varepsilon\\
E^{-3/2}\,\exp(2\sqrt{aE})&{\mathrm{for}}\ E>\varepsilon
\end{array}\right.,
\label{eq:composit}
\end{equation}
where $\eta=\varepsilon^{-3/2}\,\exp(2\sqrt{a\varepsilon}-\varepsilon/\tau)$ 
accounts for continuity at the energy $E=\varepsilon$. If we also require the 
slopes to be equal at $\varepsilon$, the level density parameter $a$ is 
restricted to
\begin{equation}
a=\left(\frac{\sqrt{\varepsilon}}{\tau}+\frac{3}{2\,\sqrt{\varepsilon}}
\right)^2.
\label{eq:condition}
\end{equation}

\begin{figure}\centering
\includegraphics[totalheight=8.9cm]{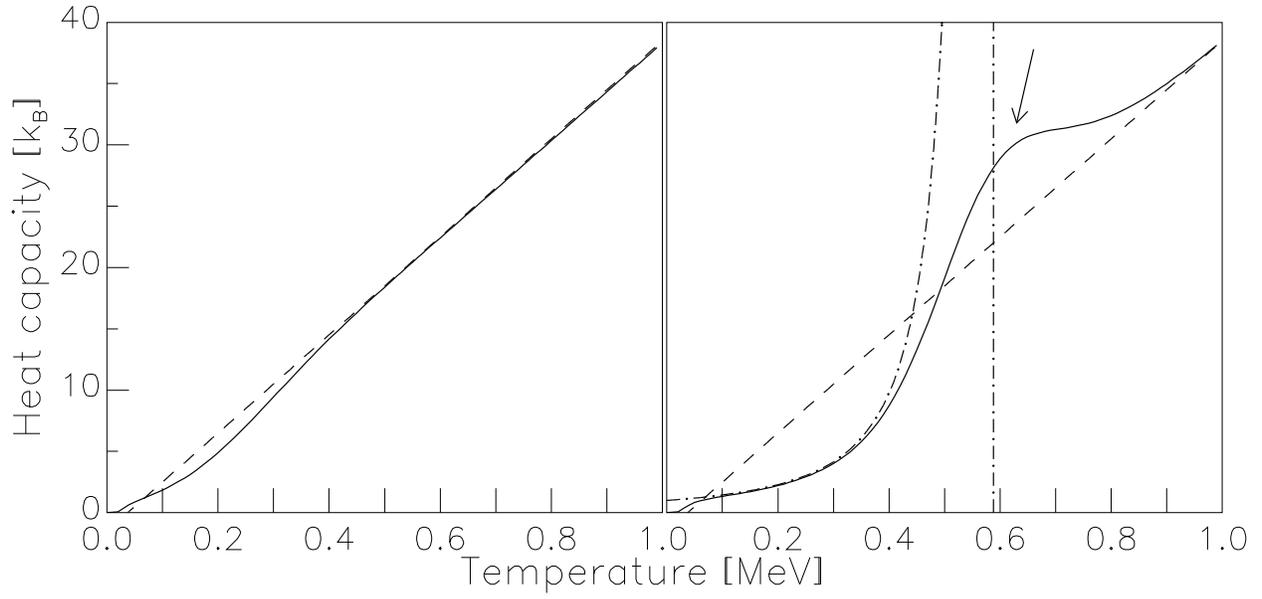}
\caption{A pure Fermi gas model cannot give rise to the characteristic S-shape
of the canonical heat capacity curve $C_V(T)$ (left panel). A simple composite
level density can simulate our experimental findings (right panel).}
\label{fig:simulation} 
\end{figure}

Figure~\ref{fig:simulation} shows the heat capacity evaluated in the canonical 
ensemble with the level density function of Eq.~(\ref{eq:composit}) and 
$\tau^{-1}=1.7$~MeV$^{-1}$. The left hand part simulates a pure Fermi gas 
description, i.e.\ the case $\varepsilon=0$, assuming a level density parameter
$a=20$~MeV$^{-1}$. One can see that a pure Fermi gas does not give rise to the 
characteristic S-shape of the heat capacity as in Fig.~\ref{fig:heatcapacity}.
The right hand part simulates the experiments, where $\varepsilon=5$~MeV and 
$a$ fulfills Eq.~(\ref{eq:condition}), i.e.\ again $a=20$~MeV$^{-1}$. The 
characteristic S-shape emerges. 

\begin{figure}\centering
\includegraphics[totalheight=17.9cm]{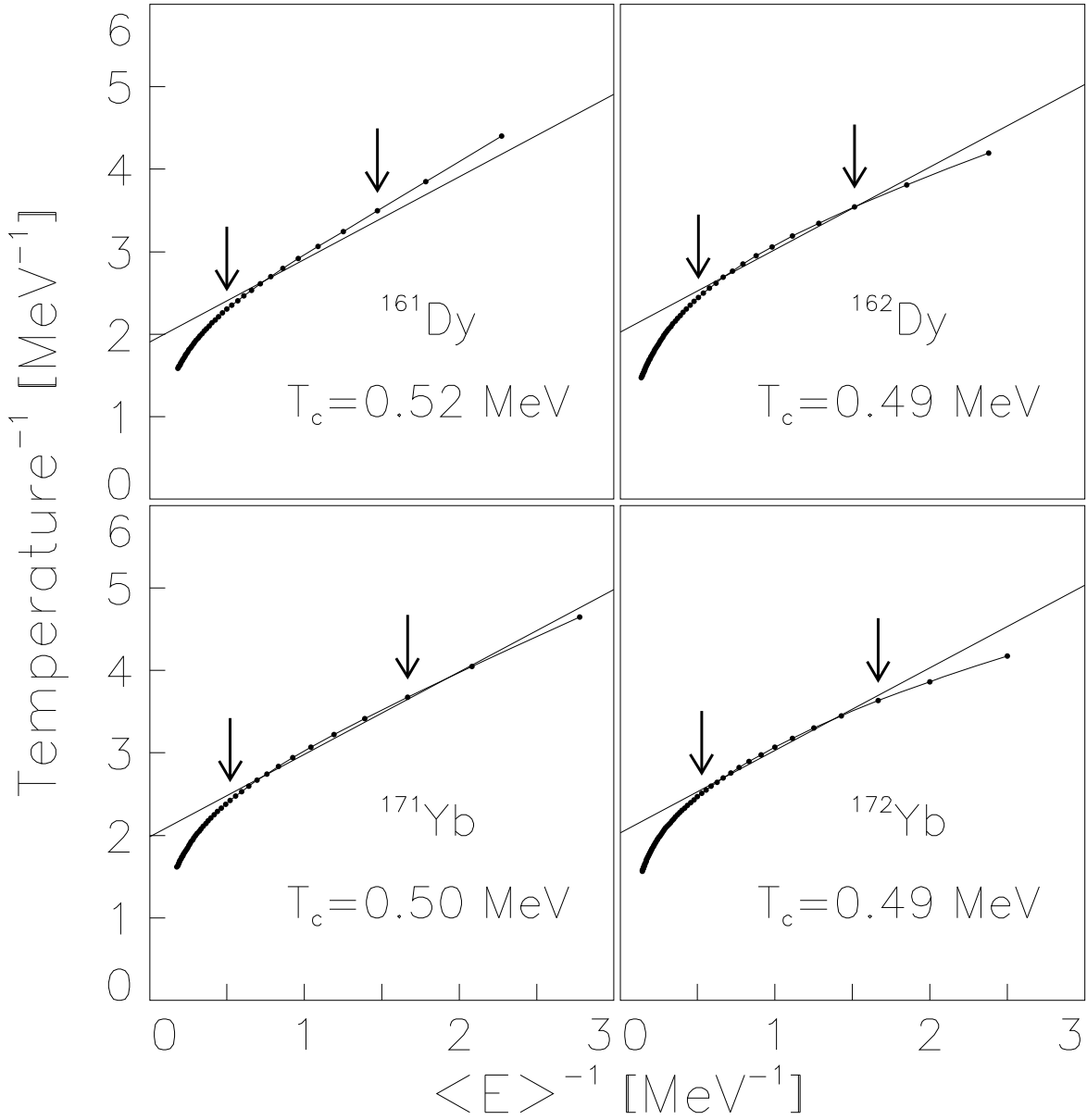}
\caption{The critical temperature is deduced by fitting a straight line with 
slope 1 to the data points between the arrows.}
\label{fig:critical} 
\end{figure}

Therefore, our method to find $T_c$ relies on the assumption that the lower 
energetic part of the level density can be approximately described by a 
constant temperature level density. Calculating $\langle E(T)\rangle$ and
$C_V(T)$ within the canonical ensemble for an exponential level density gives 
$T^{-1}=\langle E(T)\rangle^{-1}+\tau^{-1}$ and 
\begin{equation}
\label{eq:Tc}
C_V(T)=(1-T/\tau)^{-2}.
\end{equation}
Thus, plotting $T^{-1}$ as function of $\langle E(T)\rangle^{-1}$ one can  
determine $\tau$ from Fig.~\ref{fig:critical}. The quantity $\tau$ is then 
identified with the critical temperature $T_c$, since $C_V(T)$ according to 
Eq.~(\ref{eq:Tc}) exhibits a pole at $\tau$ and the analogy with the definition
of $T_\lambda$ in the theory of superfluids becomes evident. The $C_V(T)$ curve
of Eq.~(\ref{eq:Tc}) with $T_c=\tau$ using the extracted critical temperatures 
for the four nuclei is shown as dashed-dotted lines in 
Fig.~\ref{fig:heatcapacity}. This simple analytical expression with only one 
parameter $T_c$ fits the experimental data up to temperatures of $\sim$0.4~MeV.
The critical temperature itself is marked by the vertical lines. The extracted 
$T_c$ is rather close to the estimate of $T_c$ represented by the arrows.
The extracted values are $T_c=$ 0.52, 0.49, 0.50 and 0.49 MeV for the 
$^{161,162}$Dy and $^{171,172}$Yb nuclei respectively, which are somehow 
delayed compared to a degenerated BCS model with $T_c=0.5\,\Delta$ yielding 
$T_c\sim$ 0.48, 0.46, 0.41 and 0.38~MeV for the respective nuclei, where 
$\Delta$ is calculated from neutron separation energies \cite{FS96}. 

We will now discuss how sensitive the extracted critical temperature is with
respect to the extrapolation, and we will give an estimate of the uncertainty
of the extracted critical temperatures. In the fit in Fig.~\ref{fig:critical},
we use only energies from $\langle E\rangle\sim$ 0.5--2~MeV. This corresponds
to energies in the level density curves up to $E_n\sim$6~MeV according to 
Eq.~(\ref{eq:energy}). Also in Fig.~\ref{fig:heatcapacity}, the interval where
Eq.~(\ref{eq:Tc}) fits the experimental data is $T\sim$0--0.4~MeV. This
corresponds to energies in the level density curves up to $E_n\sim$8~MeV. Thus,
the extracted critical temperature does only depend weakly on the actual 
extrapolation of $\rho$ curve. Also the S-shape of the $C_V(T)$ curve depends 
only on the fact, that the nuclear level density develops somehow as a Fermi 
gas expression at energies somewhere above the neutron binding energy. However,
the actual values of the $C_V(T)$ curve above $T\sim$0.5~MeV do depend on the 
specific extrapolation chosen.

With respect to $T_c$, the extrapolation is only important for determining the 
parameters $A$ and especially $\alpha$ of Eq.~(\ref{eq:transformation}). We can
therefore estimate the error of $T_c$ by 
\begin{equation}
\left(\frac{\Delta E\,\Delta T_c}{T_c^2}\right)^2=
\left(\frac{\Delta a\,\Delta U}{2\,\sqrt{aU_n}}\right)^2+
\left(\frac{\Delta D}{D}\right)^2+2\,(0.05)^2,
\end{equation}
where $\Delta E$ is the energy difference between the upper and lower energy
where $A$ and $\alpha$ are determined. $\Delta a$ is the uncertainty of the 
level density parameter, $\Delta U$ is the energy difference between the 
neutron binding energy and the upper point where $A$ and $\alpha$ are 
determined, $U_n$ is the (shifted) neutron binding energy. $D$ and $\Delta D$ 
are the neutron resonance spacing and its error. The errors of 5\% are added in
order to account for the two fitting procedures, one fitting $A$ and $\alpha$ 
the other fitting $T_c$, both with an uncertainty of some 5\%. Using a very 
conservative estimate of $a\approx 17.5(6.0)$~MeV$^{-1}$, $U_n<8$~MeV, 
$\Delta U<3$~MeV, $D/\Delta D<0.2$ and $\Delta E>4.5$~MeV, we obtain 
$\Delta T_c<0.04$~MeV. This yields a maximum error of $T_c$ of some 8\%. It is
also important to notice, that due to the strong smoothing in 
Eq.~(\ref{eq:energy}), the errors of the experimental level density curves are
negligible in our calculation.

In conclusion, we have seen a fingerprint of a phase transition in a finite 
system for the quenching of pairing correlations as a whole, given by the 
S-shape of the canonical heat capacity curves in rare earth nuclei. For the 
first time the critical temperature $T_c$ at which pair correlations in rare 
earth nuclei are quenched, has been extracted from experimental data. The 
reminiscence of the quenching process is distributed over a 6~MeV broad 
excitation energy region, which is difficult to observe and interpret in the 
microcanonical ensemble. Simple arguments show that the peak in the heat 
capacity arises from two components in the level density; a constant 
temperature like part and a Fermi gas like part. It would be very interesting 
to compare our results with SMMC calculations performed for a narrow spin 
window.

The authors are grateful to E.A.~Olsen and J.~Wikne for providing the excellent
experimental conditions. We thank Y. Alhassid for several interesting 
discussions. We wish to acknowledge the support from the Norwegian Research 
Council (NFR).

\end{multicols}

\clearpage

\clearpage

\clearpage

\clearpage

\end{document}